\documentclass[12pt]{article}

\usepackage{amsmath}
\usepackage{amssymb}
\usepackage{amsfonts}
\usepackage{amscd}
\usepackage{amsbsy}
\usepackage{amsthm}
\usepackage{latexsym}
\usepackage{graphicx}
\usepackage{slashed}
\usepackage{color}

\def\beq{\begin{eqnarray}}
\def\eeq{\end{eqnarray}}

\def\La{\Lambda}

\begin{document}

\begin{center}

{\large\bf
Kantoswki-Sachs model with a running
\\
cosmological constant and radiation}

\vskip 5mm

\textbf{
Vin\'{\i}cius G. Oliveira,\footnote{E-mail address:
\ vinicius.guilherme@ice.ufjf.br}
\ Gil de Oliveira-Neto\footnote{E-mail address:
\ gilneto@fisica.ufjf.br}
\ and \
Ilya L. Shapiro\footnote{E-mail address:
\ ilyashapiro2003@ufjf.br}}

\vskip 5mm

{\sl Departamento de F\'{\i}sica, ICE,
Universidade Federal de Juiz de Fora
\\
Campus Universit\'{a}rio - Juiz de Fora, 36036-900, MG, Brazil}
\end{center}

\vskip 6mm

\centerline{\textbf{Abstract}}
\vskip 1mm

\begin{quotation}
\noindent
The simplest anisotropic model of the early Universe is the one with
two conformal factors, which can be identified as Kantowski-Sachs
metric, or the reduced version of the Bianchi-I metric. To fit the
existing observational data, it is important that the anisotropy is
washed out in the early stage of the evolution. We explore the
possible effect of the running cosmological constant (RCC) on
the dynamics of isotropy, in the case of the space filled by radiation.
\vskip 3mm

\vskip 2mm


\end{quotation}

\section{Introduction}
	
Two important theoretical challenges for the theoretical background
of modern cosmology are to construct the basis for a possible
variation of the equation of state of the Dark Energy and to explain
the initial conditions of the Universe. One of important aspects of
the last task is to elaborate a mechanism of making the Universe
isotropic, at least after the initial stage of its evolution which
leaves observational traces.

The most natural candidate to be Dark Energy is the cosmological
constant $\La$ (see, e.g., \cite{Weinberg89}),
which has fixed equation of state $P_\La =  - \rho_\La$
between ``pressure'' and ``energy density'' components. If the future
observational data show the deviation from this value, it maybe
either interpreted as a non-constant cosmological term or as the
presence of a qualitatively new essence filling the Universe, which
may be a replacement or a compliment to the cosmological constant.
The non-constant cosmological term may be a consequence of the vacuum
quantum effects of matter fields (see, e.g., the review \cite{PoImpo})
for a qualitative discussion and further references. The corresponding
quantum contributions to the action of gravity are certainly rather
complicated (e.g., necessarily non-polynomial) if expressed via
curvature tensors and nonlocal form factors \cite{DCCrun}. This
explains why these terms were never calculated with the existing
quantum field theory techniques based on the weak field expansions.
For the same reason, the presence of these quantum contributions
cannot
be ruled out. In this situation one can rely on the phenomenological
approaches, e.g., based on the assumption of quadratic decoupling
in the lower-derivative sector of the gravitational effective action
\cite{Babic2002,CC-nova}, or assuming and using the covariance
of the effective action \cite{Farina}. All these approaches converge
to the IR (low-energy) running of the form,
\beq
\rho_\La(\mu) \,=\,
\rho_\La^0 + \frac{3\nu}{8\pi G} \big(\mu^2 - \mu_0^2\big),
\label{runCCdens}
\eeq
where $G$ is the Newton constant and $\rho_\La^0$ is the value of the
density of the cosmological constant at the fiducial value $\mu_0$ of
the scale parameter $\mu$. The limits on the magnitude of the
phenomenological parameter $\nu$ were established in \cite{CCwave}
and \cite{CCG} in different types of the cosmological models based
on (\ref{runCCdens}). These limits were obtained by analyzing cosmic
perturbations and making comparison with the observational data. In
both cases, this analysis requires an identification of the artificial scale
parameter $\mu$ of the Minimal Subtraction renormalization scheme
with certain physical quantity, as discussed in \cite{DCCrun}.

In the cosmological setting, some physical arguments based on quantum
field theory and also the scale-setting
procedure \cite{Babic2005} hint at the identification of $\mu$
with the Hubble parameter $H$. On top of this, the covariance-based
arguments imply that, under the derivative expansion, the effective
action cannot be odd in metric derivatives. For the background
cosmological metric, this gives (\ref{runCCdens}), and the same
result follows from the assumption of quadratic IR decoupling in
the beta function of $\rho_\La$.

The IR running implies that there is an effective action of gravity
that can be separated into the nonlocal parts responsible for the
IR running of the cosmological constant, the quantum corrections
to the Einstein-Hilbert term (the running of the Newton constant),
and the terms which can be attributed to quantum corrections in the
higher derivative sectors. The last terms can be directly calculated
(see, e.g., \cite{OUP} for review and references), but are not very
relevant for the late cosmology owing to the Planck suppression of
the higher derivative terms. Thus, the covariance of the effective
action assumes that the lower-energy sector should satisfy certain
conservation laws on its own. In this respect, the cosmological
applications of (\ref{runCCdens}) can be separated in the models
admitting the energy exchange between vacuum and matter sectors
and the ones without such an exchange. It was argued in
\cite{Wang,OphPel} and \cite{CC-Gruni} that the models of the
first type are physically unappropriated for the late Universe. On
the other hand, the phenomenological limits on the parameter $\nu$
in (\ref{runCCdens}) derived from the metric perturbations and
LSS data \cite{CCwave} are much stronger in these models, as it
was also confirmed in the more recent works \cite{Agudelo2020}
by analyzing other set of cosmological observables (see also
\cite{Basilakos2019} and references therein). According to the last
works, in the early Universe
(and certainly not in the later stages of the evolution) there is
no suppression of the creation of particles from the vacuum
\cite{Wang,OphPel}, making the exchange of energy between
different parts of the gravitational action less relevant. In this
case one can use the basic cosmological models based on the
running \cite{CC-astron-2003} instead of the more complicated
models of the type considered in \cite{CC-Gruni}.

Regardless a lot of the relevant information in cosmology is obtained
from linear cosmic perturbations, there is at least one special situation
when one needs to perform a non-perturbative analysis. This concerns
the answer to the question of why the initial stage of the Universe can
be described by the isotropic metric. To address this problem, one
needs to start with the anisotropic model and see whether and how
the isotropy is restored in a given model of gravity. Since the issue
arises for the very early Universe, the matter fields can be described
by pure radiation, which is a dominating component in this epoch,
even taking into account the symmetry restoration and the
corresponding huge (compared to the present one) magnitude
of the cosmological constant \cite{BludRud,Weinberg89}.

In the present work, we report on the first (at least, up to our
knowledge) theoretical investigation of the effect of the running
of the cosmological constant density (\ref{runCCdens}) on the
isotropization of the early Universe. For this initial work we use
the simplest model including only radiation and the cosmological
constant in the gravity theory based on the Einstein's GR with the
running cosmological constant. The running of Newton constant
and other terms in the action of gravity are not taken into account,
as they are less relevant in the given physical situation in the
early Universe, when the energy exchange between vacuum and
matter sectors of the action are not suppressed \cite{Agudelo2020}.
Finally, to explore the anisotropy we use the simplest version of
the Bianchi type I metric, which is also some version of the
Kantowski-Sachs (KS)  model  \cite{KSori}.
This metric has only two conformal
factors and enables one to explore the main qualitative features
of the anisotropic running cosmology in the most economic and
explicit way. It is worth noting that isotropization in the KS
cosmological models without running was previously explored
in many papers, including \cite{HeckSchuck} and \cite{Jacobs},
where the isotropization of the metric was first discovered
(see also \cite{weber,gron,barrow,byland,adhav,parisi,gil} for
further investigations in different models and
\cite{Khalatnikov2003} and \cite{Casadio2023} for a more
complete set of references).  It is worth noting the quantum 
mechanism of isotropization (see, e.g., 
\cite{Zeldovich1971,Collins1972,Kofman1983}, 
there are also many other papers on this issue, and the review 
in the book \cite{GMM-94}). 

The rest of the work is organized as follows. In the next
Sec.~\ref{sec2} we formulate the background for the anisotropic
running cosmology, that includes the identification of scale and
derivation of the main formulas for the dynamics of the conformal
factors. Let us note that the generalization to more complicated
metrics, such as the general Bianchi-I, is expected straightforward.
Sec.~\ref{sec3} reports on the numerical results for the dynamics
of the conformal factors. Finally, in \ref{secConc} we draw our first
conclusions and discuss possible extensions of the present work.

\section{Theoretical background of the anisotropic running cosmology}
\label{sec2}

The basis of our investigation will be Einstein's equations with
the cosmological constant,
\beq
G_{\alpha \beta}\,=\,
8\pi G \,T_{\alpha \beta} + \La g_{\alpha \beta},
\label{eq einstein}
\eeq
where $g_{\alpha \beta}$ is the metric tensor, the Newton
constant $G$ is assumed to be scale-independent, as explained
above, and $\La = 8\pi G \rho_\La$ depends on the scale
parameter $\mu$ according to (\ref{runCCdens}). Here and
in what follows we adopt the units with $c = 1$ for the speed
of light in vacuum.
	
Consider the Kantowski-Sachs metric
\beq
ds^{2}\,=\,
-\, dt^2 + a^2(t)dr^2
+ b^2(t) \big[ d\theta^{2} + \sin^2 \theta d\phi^2\big],
\label{metrica}
\eeq
where $r$, $\theta$ and $\phi$ are spherical coordinates, $a(t)$ and
$b(t)$ are the two scale factors. The growth of these functions with
time characterizes the expansion of the Universe. In the model
(\ref{metrica}), the radial part can expand differently than the
angular parts. Since there are only two functions, this is one of
the simplest possible anisotropic models. The spatial sections of
this model have positive curvature.
		
The energy-momentum tensor for the perfect fluid is given by
\beq
T_{\alpha \beta}
\,=\,\big(\rho_f + p_f\big)u_{\alpha}u_{\beta} + pg_{\alpha \beta},
\label{Tenergiamomento}
\eeq
where $\rho_f$ and $p_f$ are, respectively, the energy density and
pressure of the fluid and $u_{\alpha}$ is the four-velocity of the
fluid. Since we are interested in the very early Universe, the matter
contents may be approximately described by radiation, so the
equation of state for our perfect fluid should be
\beq
p_f \,=\,\frac{\rho_f}{3}\,.
\label{equationofstate}
\eeq

One may identify the isotropization of metric (\ref{metrica}) in two
different ways. In a more simple way, after some time $a(t)$ would
tend to $b(t)$. The second way is to see that the ratio between the
scale factors tends to a constant after some time, showing that the
scale factors would have the same expansion rate.

Using the KS metric (\ref{metrica}) in the Einstein tensor in the
\textit{l.h.s.} of (\ref{eq einstein}), we arrive at the system of three
ordinary differential equations,
\beq
&&
\frac{2\dot{a}\dot{b}}{ab}
+ \frac{\dot{b}^{2}}{b^2}
+ \frac{1}{b^2}
\,=\, 8\pi G \rho_t ,
\label{EqEt}
\\
&&
2b\ddot{b}+\dot{b}^2 + 1
\,=\, -\, 8\pi G  b^2\,p_t ,
\label{EqEr}
\\			
&&
\frac{\ddot{a}}{a} + \frac{\dot{a}\dot{b}}{ab}
+ \frac{\ddot{b}}{b}\,=\,
-\,8\pi G p_t.
\label{EqEang}
\eeq
In these equations $\rho_t$ and $p_t$ are the total energy density
and pressure, as it will be specialized below.

Since there are only two variables $a(t)$ and $b(t)$, we can restrict
the consideration by Eq.~(\ref{EqEt}) and the difference between
(\ref{EqEang}) multiplied by $ab^2$ and (\ref{EqEr}) multiplied by
$a$. Thus, the equations which we will work with are\footnote{We
have checked that any couple of the three equations (\ref{EqEt}),
(\ref{EqEr}) and (\ref{EqEang}) provide the same solutions.}
\beq
&&
2b\dot{a}\dot{b}+a\dot{b}^2+a
\,=\,8\pi G ab^2 \rho_t,
\label{Friedmann1}
\\
&&
b^2\ddot{a}-ab\ddot{b}+b\dot{a}\dot{b}-a\dot{b}^2-a\,=\,0.
\label{Friedmann2}
\eeq

To simplify notations, in what follows we use the units with
$8\pi G/3=1$. Together with $c = 1$, this means physical time
$t$ is measured in the Planck units. This is certainly a very small
unit, but for the very early Universe this may be a useful choice.
Concerning the \textit{r.h.s.} of Eq.~(\ref{eq einstein}), we meet
the sum of the radiation and the contribution of the variable
cosmological constant (\ref{runCCdens}). A useful representation
is by using ``energy density'' and
``pressure'' of the vacuum. Then, we may arrive at the total energy
density and pressure of the model, in the forms
\beq
\label{rho_T}
\rho_t \,=\,  \rho_f + \rho_\La,
\qquad
p_t \,=\, \frac13\,  \rho_f - \rho_\La,
\eeq
where we used the relation (\ref{equationofstate})
for the radiation and the relation $p_\La \, = \, - \rho_\La$. Let us
note that this relation between ``energy density'' and ``pressure''
of the vacuum correspond to the natural separation of the
effective action of vacuum into cosmological constant sector,
Einstein-Hilbert sector and higher-derivative part. In the isotropic
metric case, this separation, which was already mentioned in the
Introduction, can be performed using the global scaling. The
cosmological constant and the corresponding nonlocal quantum
corrections should have the same scaling and
this means the equation of state $p_\La\,=\, - \rho_\La$. The
interested reader may find more details in \cite{RadiAna}.

The next problem is an identification of $\mu$, that would enable
us to use the result in (\ref{runCCdens}) and then in (\ref{rho_T}).
We shall use the usual choice of $\mu \sim H$, and the definition
of an average $H$ suggested in \cite{adhav},
\beq
H \,=\,
\frac13\,\Big(\frac{\dot{a}}{a}\,+\,2\,\frac{\dot{b}}{b}\Big).
\label{hubble}
\eeq
This choice has several advantages. In the QFT framework,
one has to deal with the Feynman diagrams with external
gravitational lines and presume that in the cosmological
constant sector there is a quadratic decoupling. In our case,
there may be lines corresponding to different conformal
factors. In case the magnitudes in the two terms are of the
same order, it boils down to the usual identification from
\cite{Babic2002,CC-nova,Babic2005}. On the other hand,
if the ratios $\dot{a}/a$ and $\dot{b}/b$ are very different,
the choice (\ref{hubble}) guarantees that the larger version
of Hubble parameter gives greater contribution, as it has to be.
From the phenomenological side, this definition looks
natural and enables one to implement the running (\ref{runCCdens})
in the anisotropic setting. The generalization to the Bianchi-I model
is straightforward.

The energy conservation condition gives the equation
\beq
\label{energyconservation}
\dot{\rho_f} + \dot{\rho_\La}
\,+\, 3H(p_f + \rho_f+ p_{\Lambda}+ \rho_{\Lambda}) \,=\,0.
\eeq
In the units we use, the running corresponds to the relation
\beq
\rho_{\Lambda}=\rho_{\Lambda}^{0} + \nu (H^{2}-H_{0}^2).
\label{runCC-H}
\eeq
Taking in account Eq.~(\ref{equationofstate}), the equation of
state for the cosmological constant and the definition
of Hubble parameter (\ref{hubble}), after some calculations we find
the following energy conservation equation (\ref{energyconservation}),
\beq
&&
\dot{\rho_f} a^3 b^3 \,+\, \frac{4}{3}\rho_f
\big( \dot{a}a^{2}b^{3} + 2\dot{b}a^3b^2\big)
\,+\, \frac{2\nu}{9}\Big[\dot{a}\ddot{a}ab^3
\,-\, \dot{a}^{3}b^{3}-4a^{3}\dot{b}^3
\nonumber
\\
&&
\qquad
+ \,\, 4a^{3}b\dot{b}\ddot{b}
+ 2a^{2}b^{2}(\dot{a}\ddot{b}
+ \ddot{a}\dot{b})
- 2a^{2}b\dot{a}\dot{b}^2
- 2ab^{2}\dot{a}^{2}\dot{b}\Big]\,=\,0.
\label{conserv}
\eeq
For the total energy density of $\rho_t$
and using Eq.~(\ref{Friedmann1}), we arrive at the equation
\beq
(18-4\nu)ab\dot{a}\dot{b}
+ (9-4\nu)a^2\dot{b}^2
- \nu b^2\dot{a}^2+9a^2
\,=\,
9a^2b^2 \big(\rho_f +\rho^0_\La - \nu H_0^2\big).
\label{FriedmannI}
\eeq
It is worth noting that here $\rho_f$ is the energy density of
radiation which is one of the variables that has the dynamics
to be defined from the equations and $\rho^0_\La$ is the
initial point of the renormalization group flow.

\section{Numerical results for the anisotropic metric}
\label{sec3}

Solving the system of equations (\ref{Friedmann2}), (\ref{conserv})
and (\ref{FriedmannI}), one can explore the dynamics of the relevant
functions $a(t)$, $b(t)$ and $\rho_f(t)$. Let is report on the
corresponding numerical analysis.
		
Different from the previous work \cite{CCwave},
we do not consider cosmic perturbations, however the background
geometry is more complicated owing to anisotropy. On the other
hand, we know that the metric in the Universe filled by radiation
becomes isotropic in a very short time, hence our interest concerns
very early Universe. In this case, the limitations on the sign and
magnitude of the parameter $\nu$, which were established in
\cite{CCwave} (also in \cite{CCG} for another model with running
cosmological constant), do not apply anymore and we hence can
assume much greater values of $\nu$, both positive and negative.
Following this logic, we studied different cases, varying
the values of the parameters including $\nu$, in the first place.
One of our targets is the isotropization of metric (\ref{metrica}),
 i.e., evaluation of the ratio $b/a$.

The results of numerical analysis can be seen in the figures.
Let us first summarize the general features of different models,
characterized by different values of the parameter $\nu$ and
different initial data. We found that for the physically relevant
solutions, i.e., when the value of $\nu$ is small, both $a(t)$ and
$b(t)$ always expand and that $\rho_f(t)$ always tends to zero,
starting from a given initial value.

\begin{figure}
\begin{minipage}[t]{0.49\textwidth}
	\centering
\includegraphics[width=\linewidth]{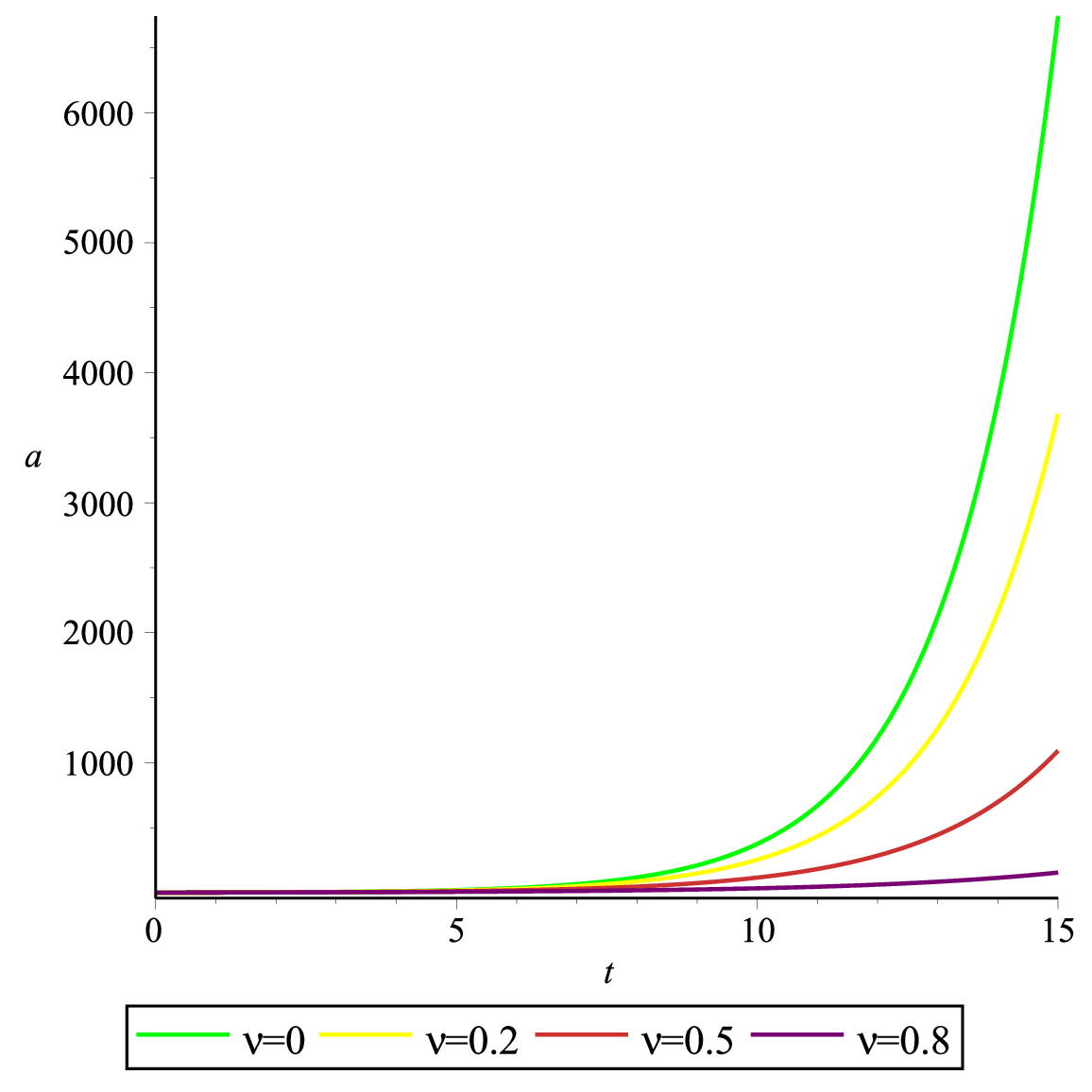}
	\caption{Variation of $a(t)$ for four different positive values of $\nu$.}
\label{f1}
   \end{minipage}\hfill
  \begin{minipage}[t]{0.49\textwidth}
\centering
\includegraphics[width=\linewidth]{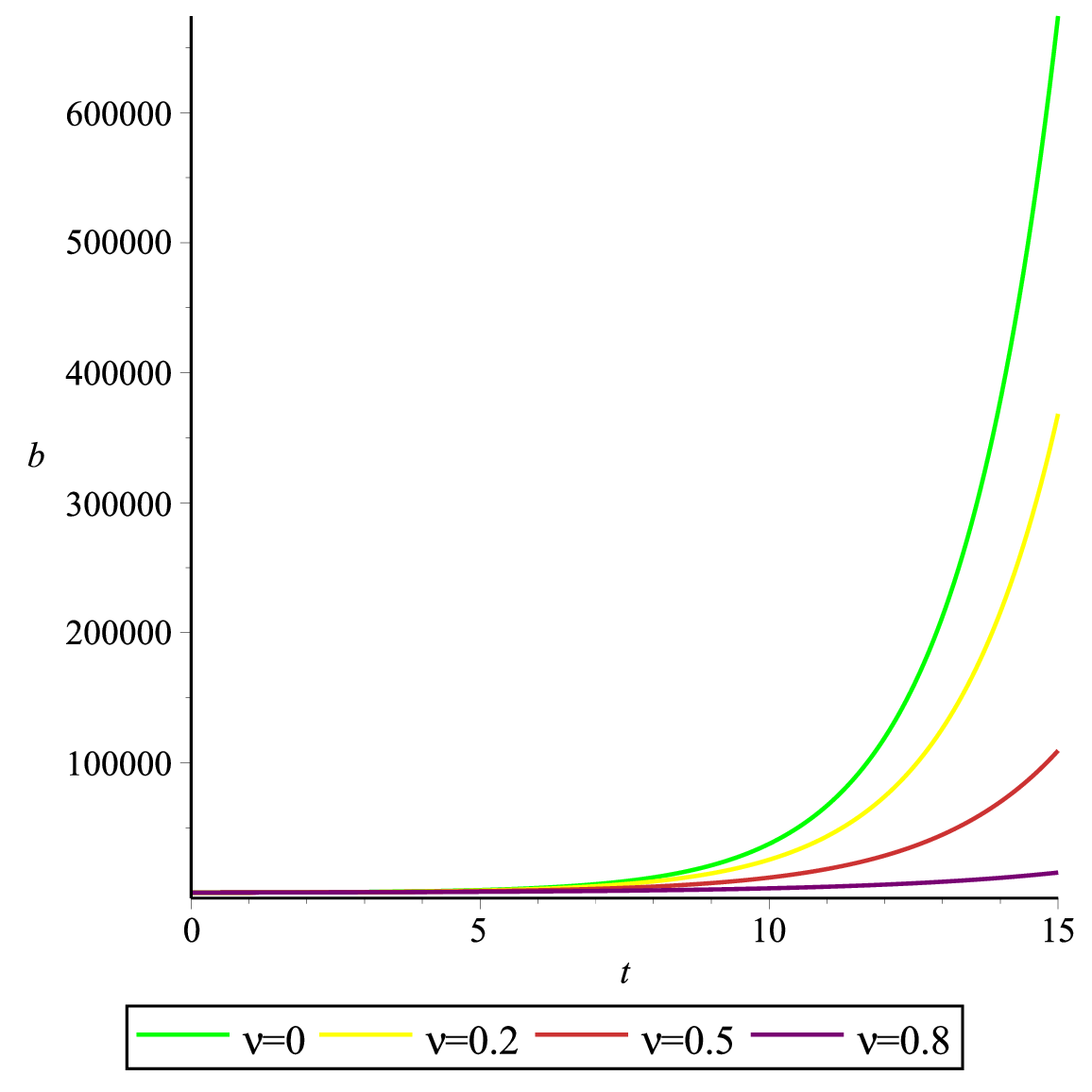}
	\caption{Variation of $b(t)$ for four different positive values of $\nu$.}
\label{f2}
\end{minipage}\hfill
\end{figure}

Consider the case when initially the model is strongly
anisotropic, that is, we choose $b(t=0)=100$ and $a(t=0)=1$.
For the numerical analysis, we used the initial values
\beq
\rho_{\Lambda}^{0} = H_0 = 1,
\qquad
\dot{a}(t=0) = 1,
\qquad
\rho_f(t=0) = 2,
\label{Larho}
\eeq
while the value of $\dot{b}(t=0)$ was varying.
In Fig.~\ref{f1} and Fig.~\ref{f2}, we show some plots obtained
by the variation of $\nu$. One can see both $a(t)$ and $b(t)$ are
rapidly growing with time, and it looks like the anisotropy does not
change significantly, for all values of $\nu$. Let us note that we
took much greater values of $|\nu|$ compared to the upper bounds
derived in \cite{CCwave,CCG}. There are two reasons for doing so.
The first one is that for the values of the order $10^{-6}$, which
are typical for the models of the first type (with the exchange of
energy between vacuum and matter \cite{CCwave}), the plots are
not visually distinguishable from the one for $\nu=0$. The second
reason is that the isotropization occurs very fast, when the values
of Hubble parameter are huge. Obviously, this makes sense only
assuming that the isotropization  takes place in the very early
Universe, where typical energies are very high. This means, there
is no decoupling of the highest-mass particles, providing small
values of  $\nu$ \cite{CC-nova,Babic2002}
and, therefore, there is no contradiction in assuming the
values of order one. The same concerns the sign, which was
advocated positive in \cite{CCwave}. In the effective
decoupling-based framework formulated in \cite{CC-nova}
(see also \cite{DCCrun,PoImpo}) this sign is defined by the
spin of the highest-mass particles in the spectrum beyond the
Minimal Standard Model. And if the scale of decoupling
dramatically changes, we have to take into account the
possibility of the fermion domination and, therefore, consider
also the negative values of $\nu$.

\begin{figure}
\begin{minipage}[t]{0.49\textwidth}
	\centering
	\includegraphics[width=\linewidth]{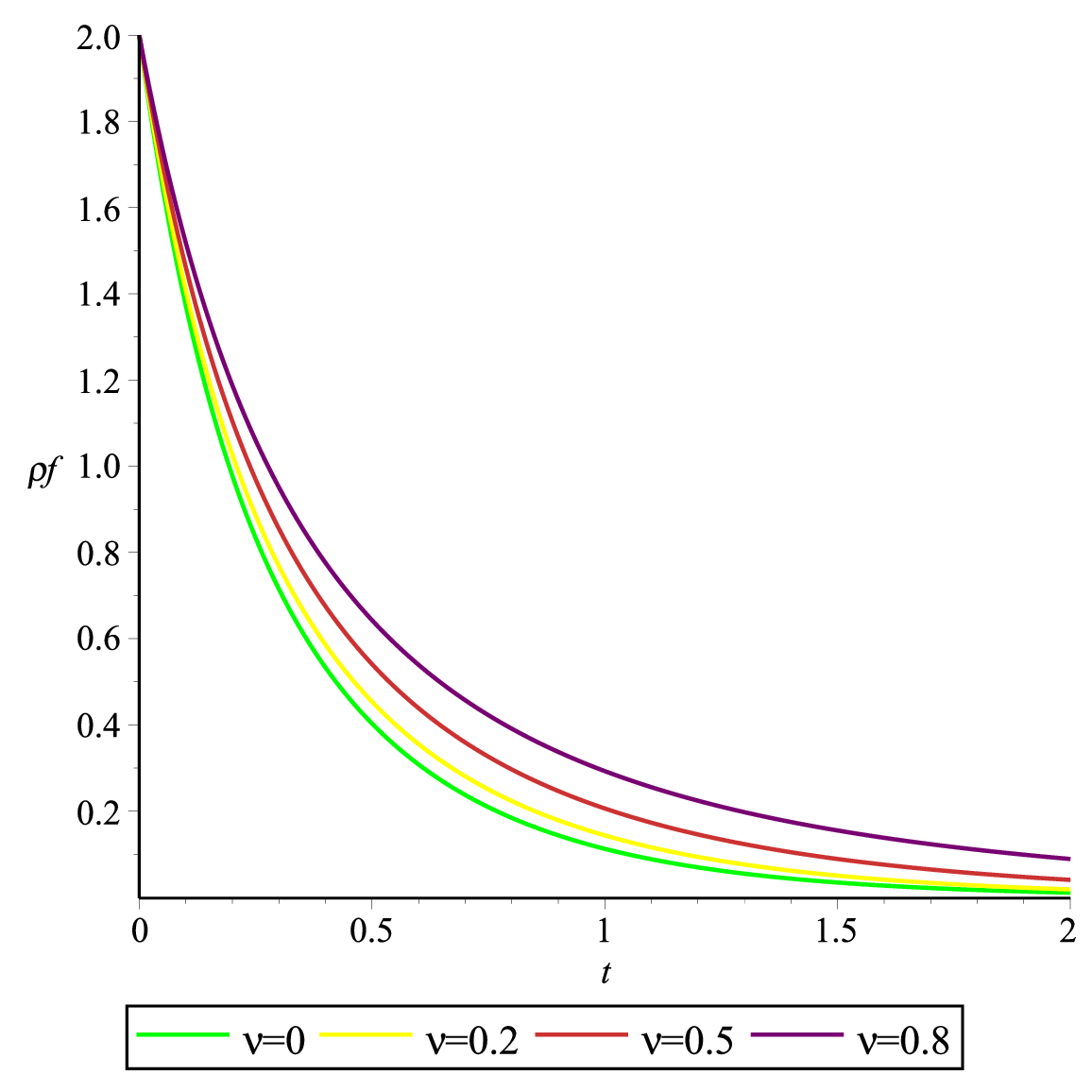}
	\caption{Variation of $\rho_f (t)$ for four different positive values of $\nu$.}
\label{f3}
	\end{minipage}\hfill
	\begin{minipage}[t]{0.49\textwidth}
\centering
	\includegraphics[width=\linewidth]{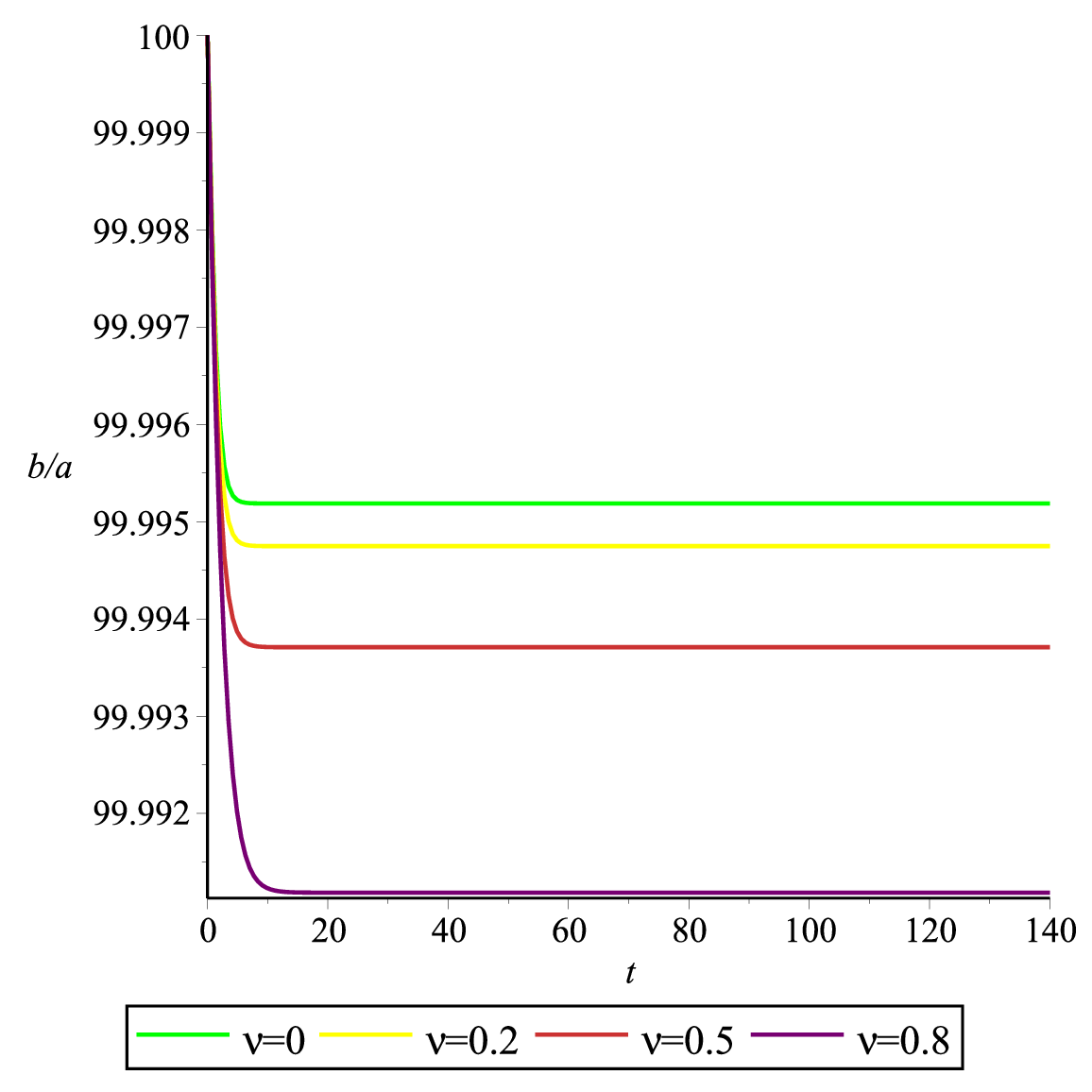}
	\caption{Variation of $b(t)/a(t)$ for four different positive values of $\nu$.}
	\label{f4}
\end{minipage}\hfill
\end{figure}

\begin{figure}
\begin{minipage}[t]{0.49\textwidth}
	\centering
\includegraphics[width=\linewidth]{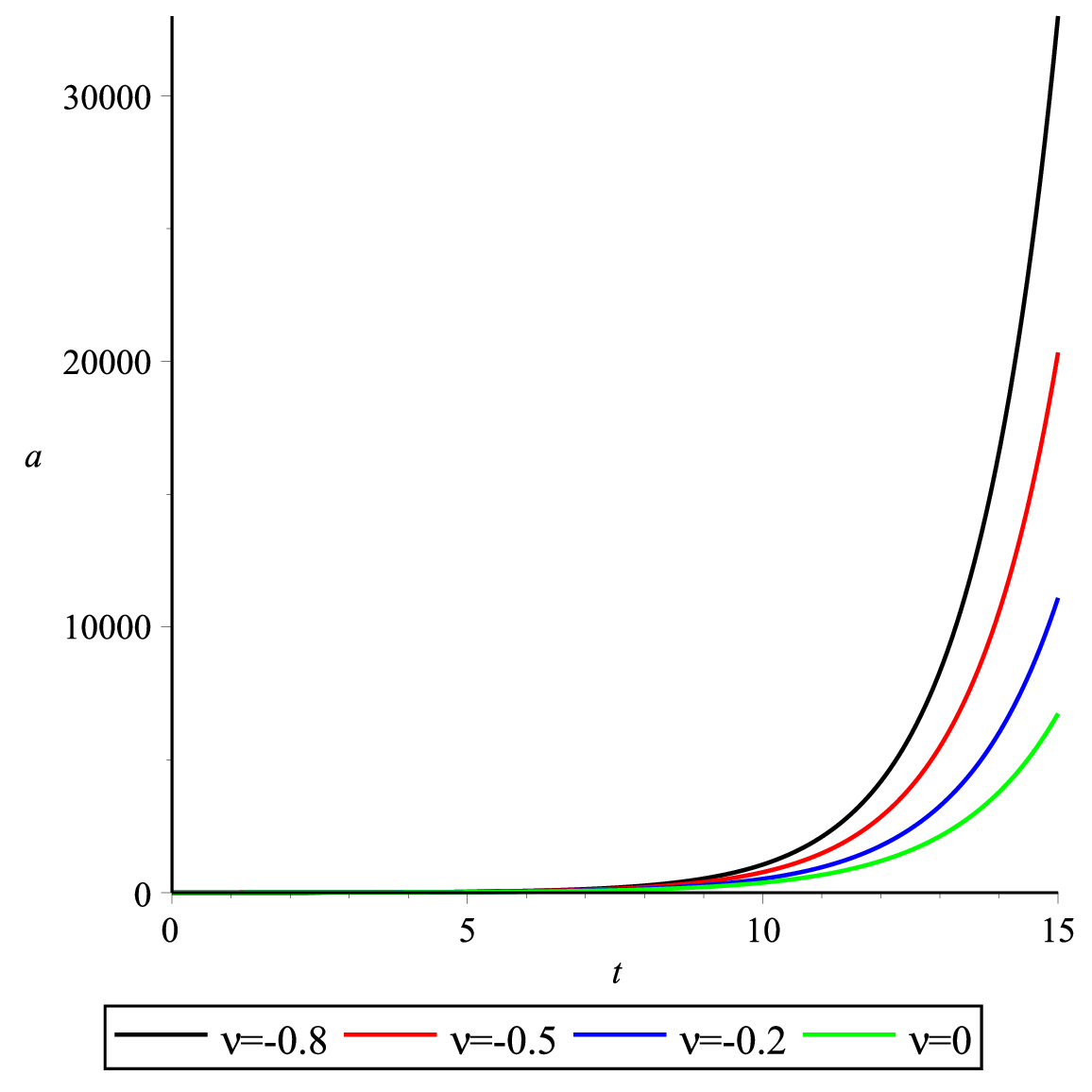}
	\caption{Variation of $a(t)$ for four different negative values of $\nu$.}
\label{f5}
\end{minipage}\hfill
\begin{minipage}[t]{0.49\textwidth}
		\centering
		\includegraphics[width=\linewidth]{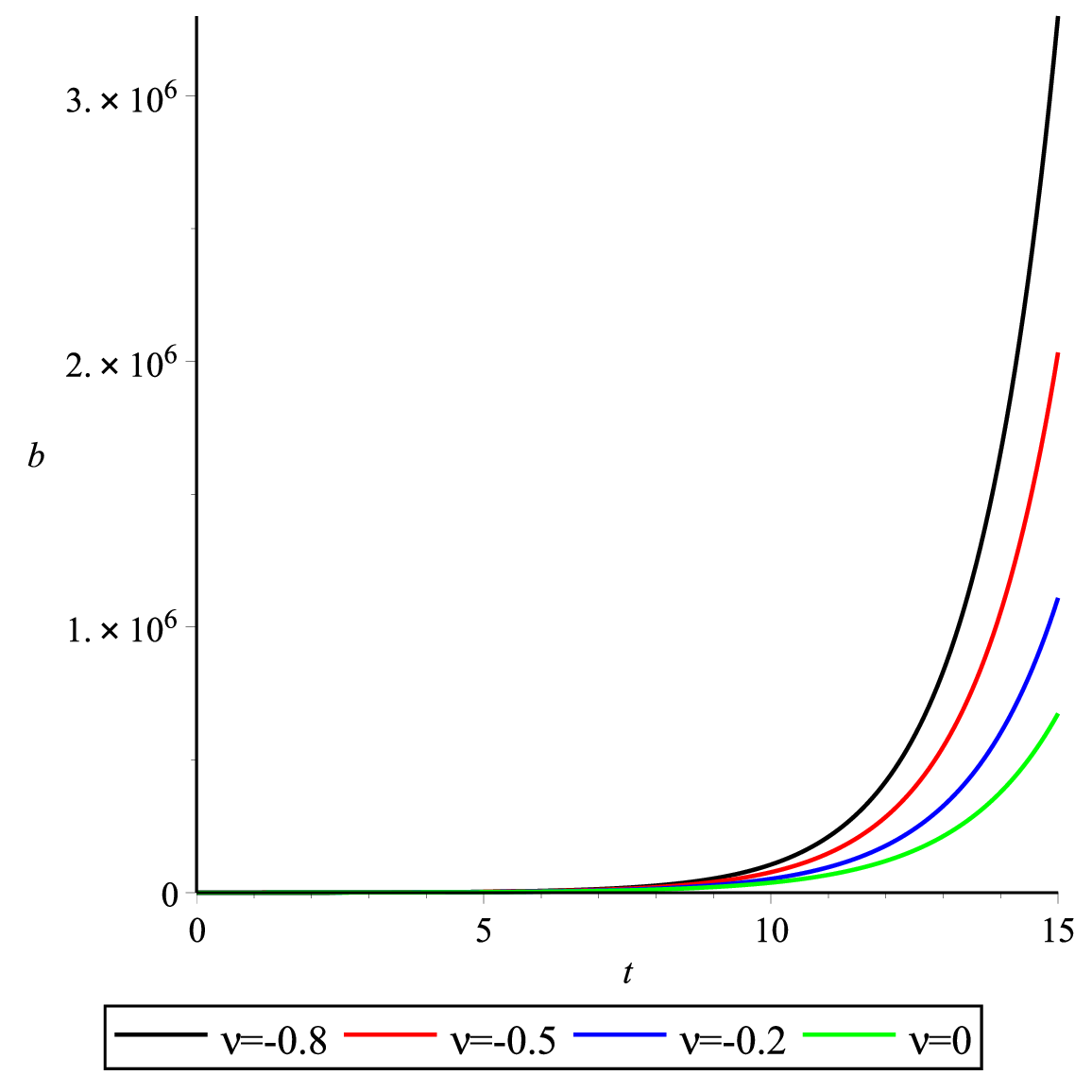}
		\caption{Variation of $b(t)$ for four different negative values of $\nu$.}
\label{f6}
\end{minipage}\hfill
\end{figure}

\begin{figure}
\begin{minipage}[t]{0.49\textwidth}
				\centering
	\includegraphics[width=\linewidth]{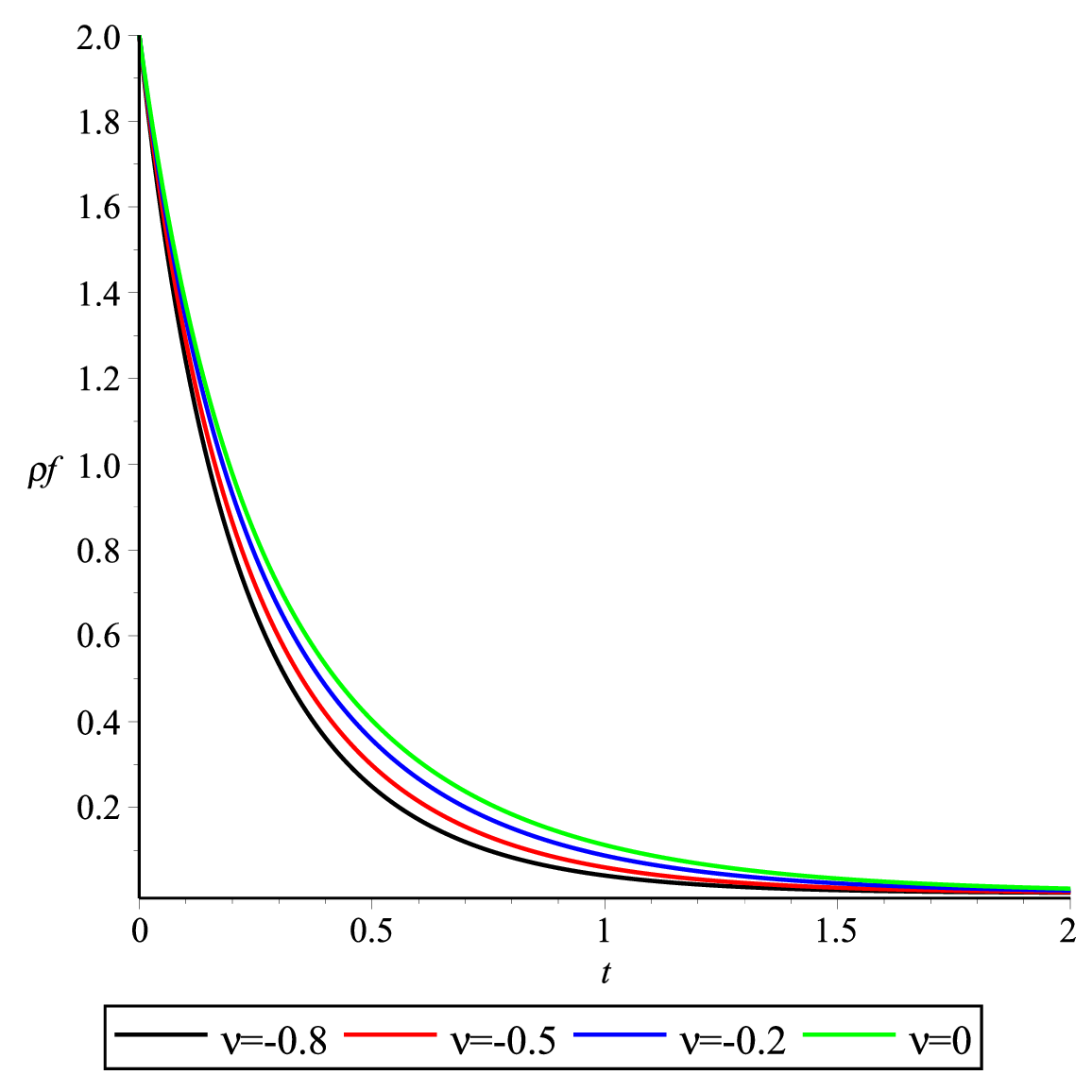}
	\caption{Variation of $\rho_f (t)$ for four different negative values of $\nu$.}
				\label{f7}
\end{minipage}\hfill
\begin{minipage}[t]{0.49\textwidth}
	\centering
	\includegraphics[width=\linewidth]{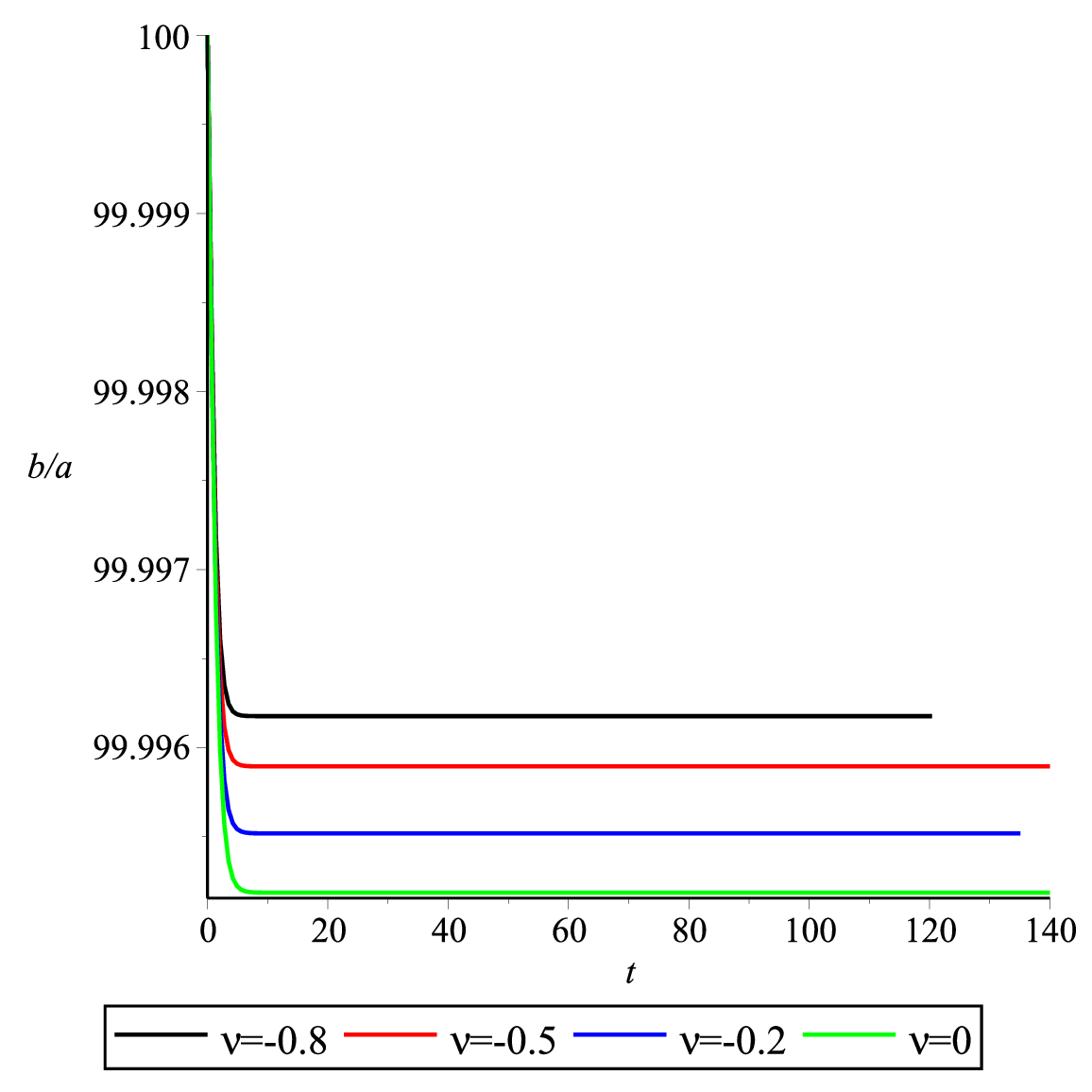}
	\caption{Variation of $b(t)/a(t)$ for four different negative values of $\nu$.}
\label{f8}
\end{minipage}\hfill
\end{figure}

\begin{figure}
\begin{minipage}[t]{0.49\textwidth}
	\centering
	\includegraphics[width=\linewidth]{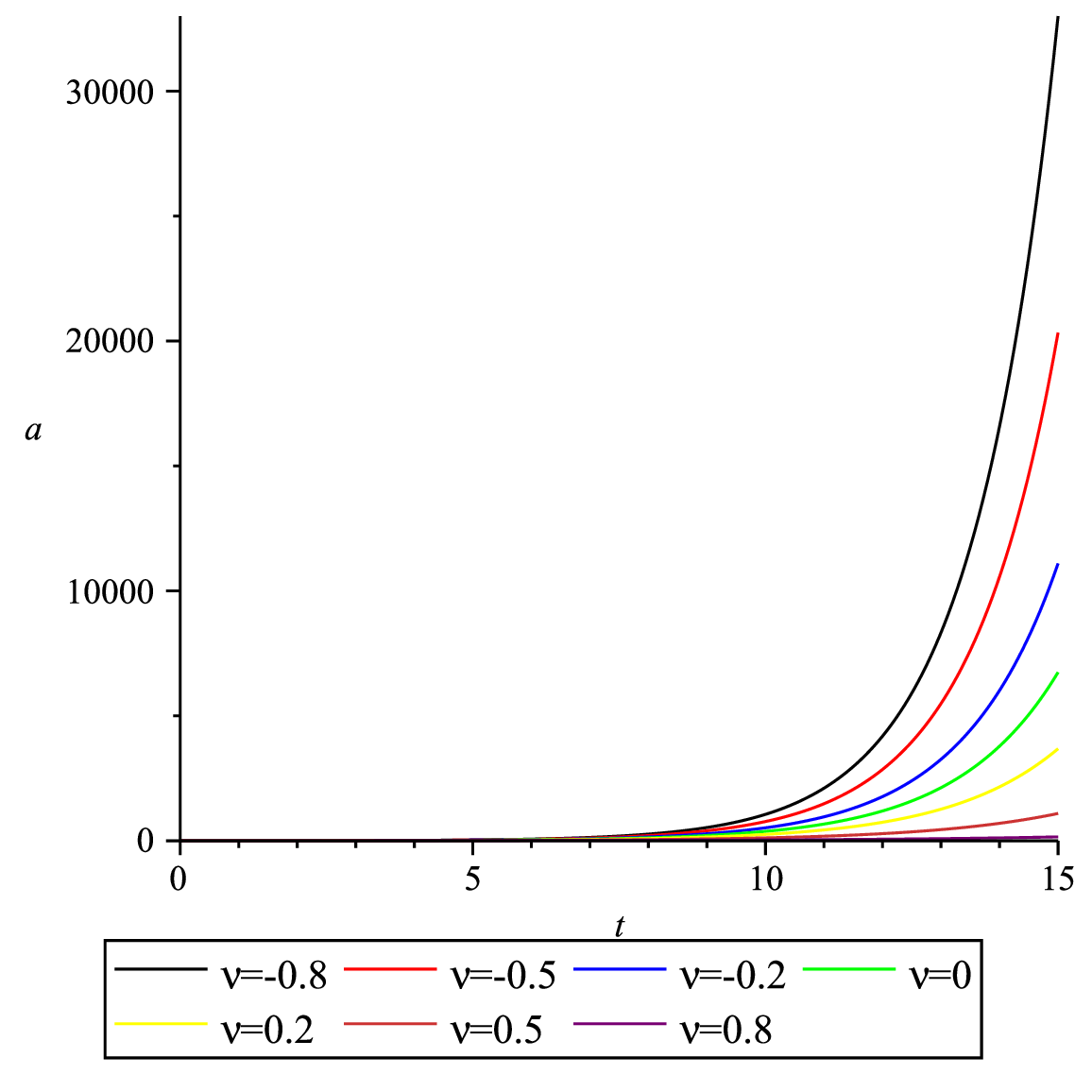}
	\caption{Variation of $a(t)$ for seven different positive and
negative values of $\nu$.}
\label{f9}
   	\end{minipage}\hfill
	\begin{minipage}[t]{0.49\textwidth}
	\centering
   	\includegraphics[width=\linewidth]{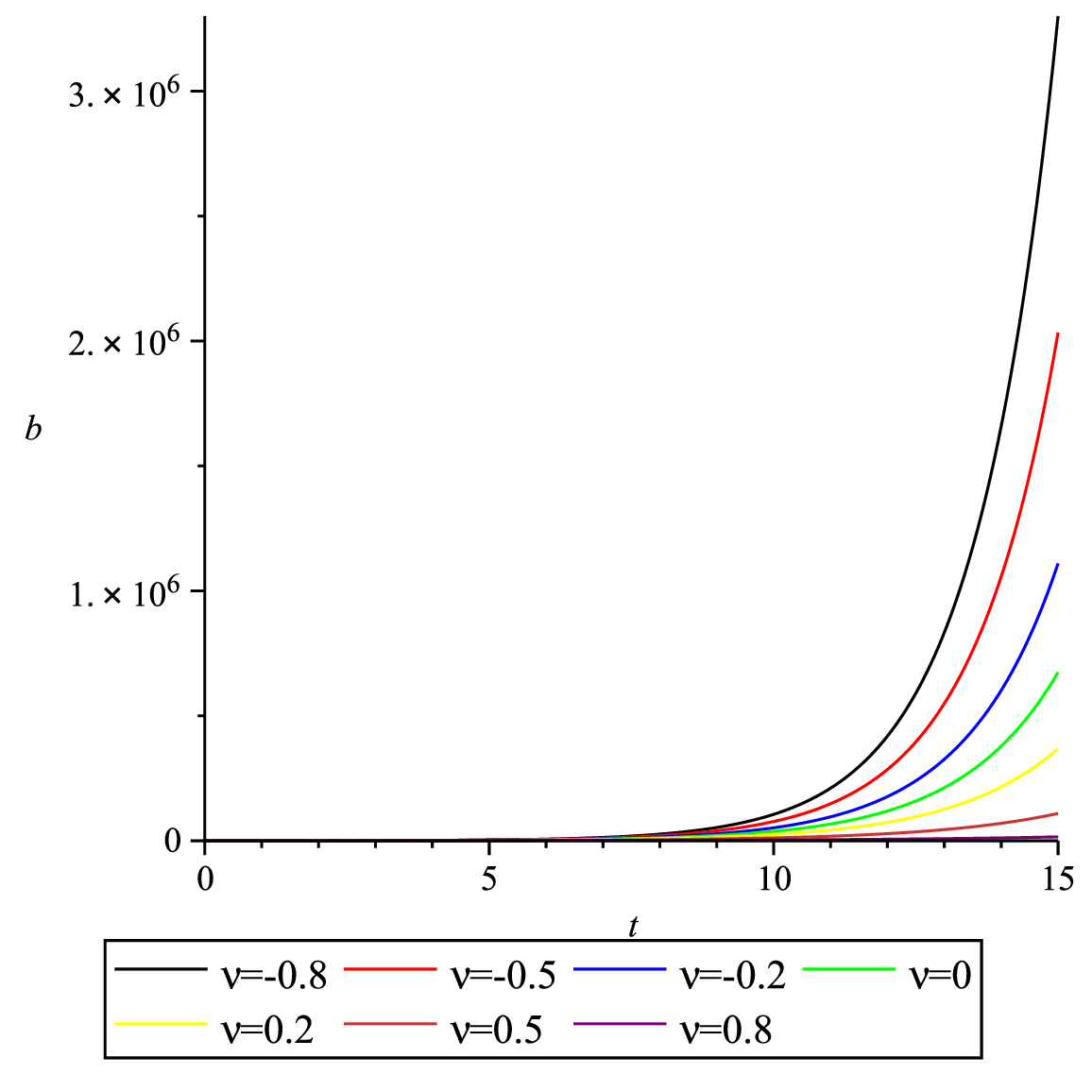}
	\caption{Variation of $b(t)$ for seven different positive and negative
values of $\nu$.}
	\label{f10}
\end{minipage}\hfill
\end{figure}

\begin{figure}
\begin{minipage}[t]{0.49\textwidth}
		\centering
		\includegraphics[width=\linewidth]{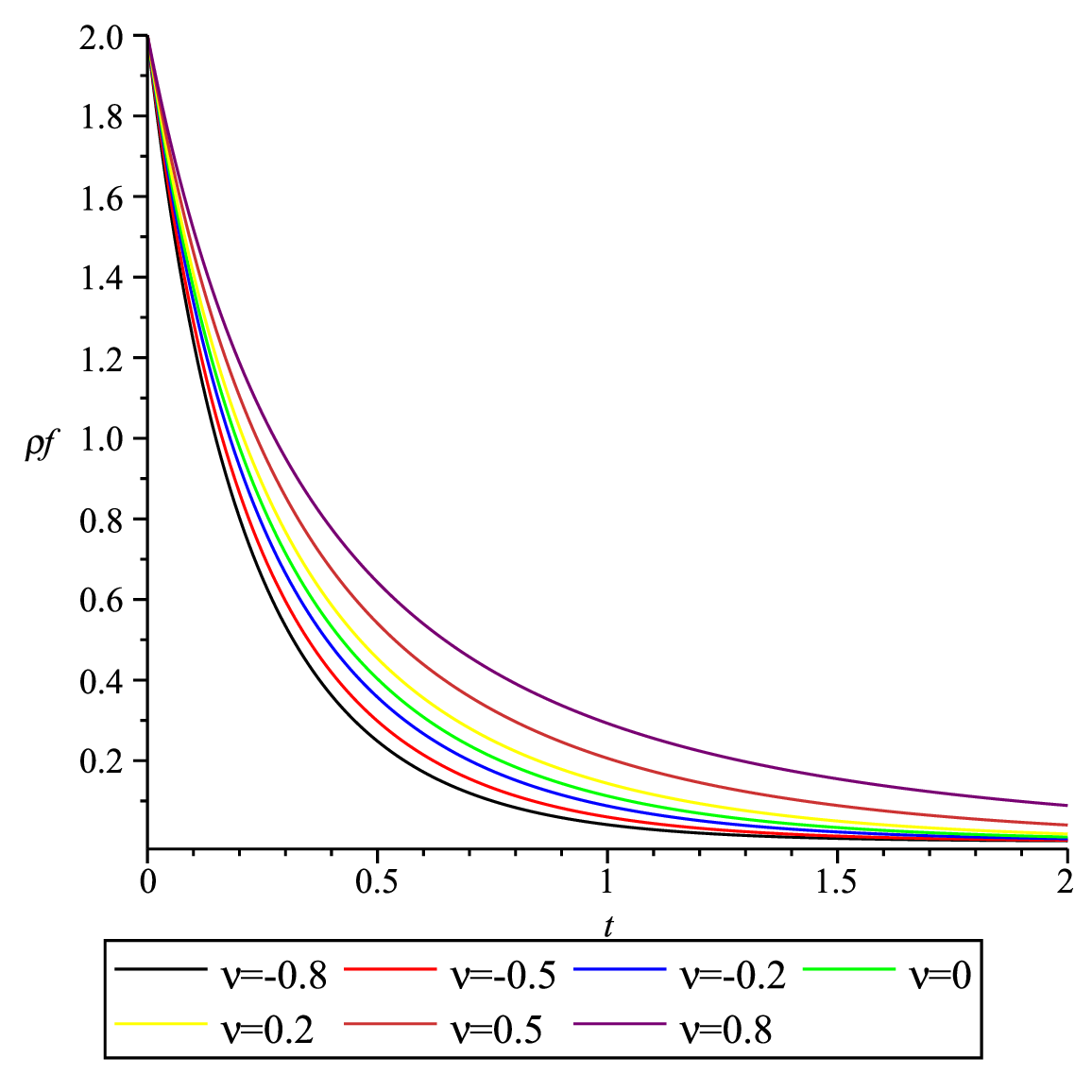}
		\caption{Variation of $\rho_f (t)$ for seven different positive and
                      negative values of $\nu$.}
\label{f11}
\end{minipage}\hfill
\begin{minipage}[t]{0.49\textwidth}
		\centering
    	\includegraphics[width=\linewidth]{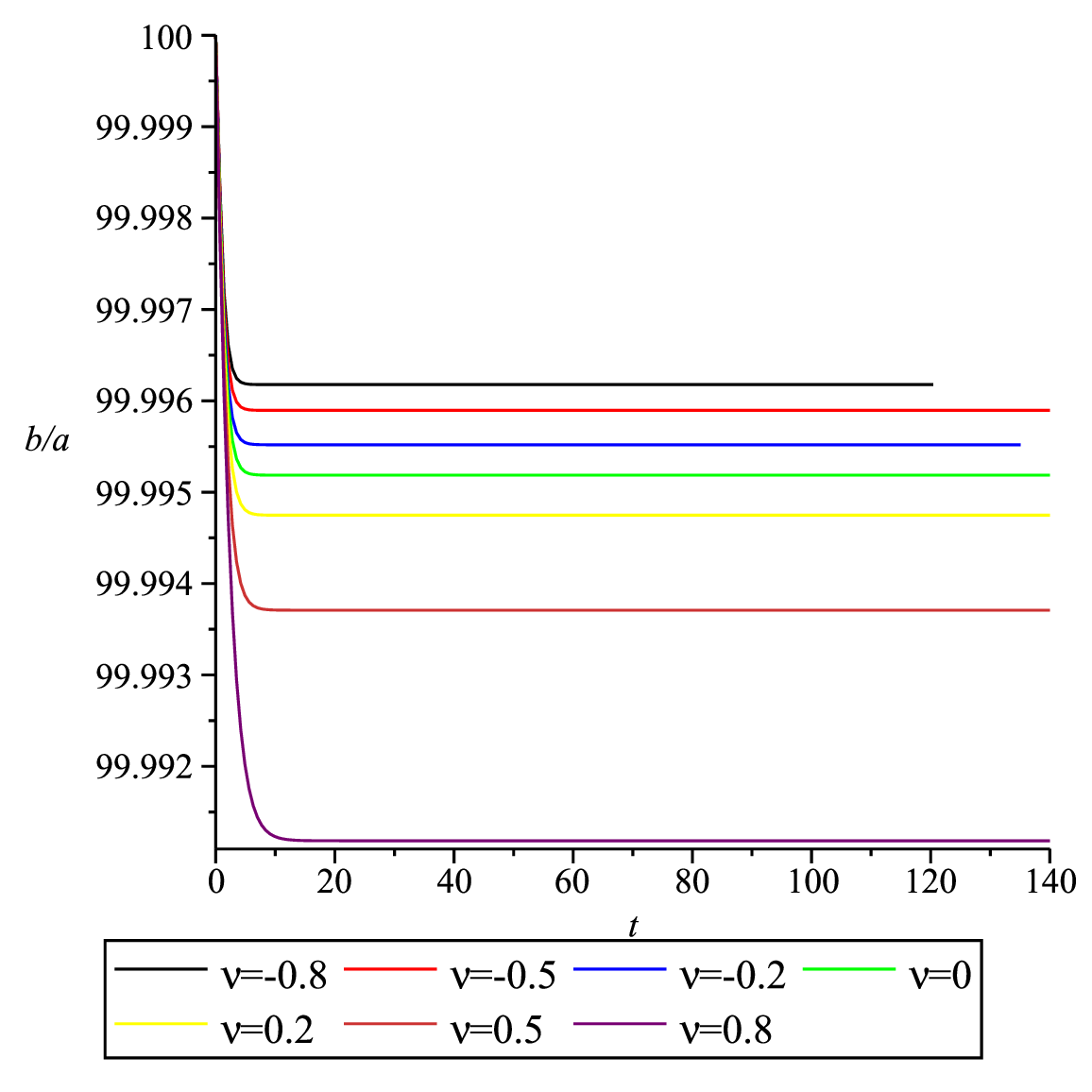}
		\caption{Variation of $b(t)/a(t)$ for seven different positive and
negative values of $\nu$.}
\label{f12}
\end{minipage}\hfill
\end{figure}

Following these arguments we choose the values for the parameters,
initial conditions and the values of $\nu$ to produce the graphs
demonstrating qualitative properties of the solutions.
Figs. \ref{f1}, \ref{f2}, \ref{f3} and \ref{f4} show, respectively,
the time dependencies $a(t)$, $b(t)$, $\rho_f (t)$ and the ratio
$b(t) / a(t)$ for four different large positive values of $\nu$.
Similarly, Figs. \ref{f5}, \ref{f6}, \ref{f7} and \ref{f8} show the
time variations of the same quantities $a(t)$, $b(t)$, $\rho_f (t)$
and $b(t) / a(t)$ for four different negative values of $\nu$.
Furthermore, Figs. \ref{f9}, \ref{f10}, \ref{f11} and \ref{f12}
illustrate the behavior of $a(t)$, $b(t)$, $\rho_f (t)$
and $b(t) / a(t)$ for four different positive and negative values
of $\nu$. Observing these plots we can see the general situation,
i.e., how the running of the cosmological constant density may
affect the process of isotropization. These general features are
formulated in the next section.

\begin{figure}
\begin{minipage}[t]{0.49\textwidth}
	\centering
	\includegraphics[width=\linewidth]{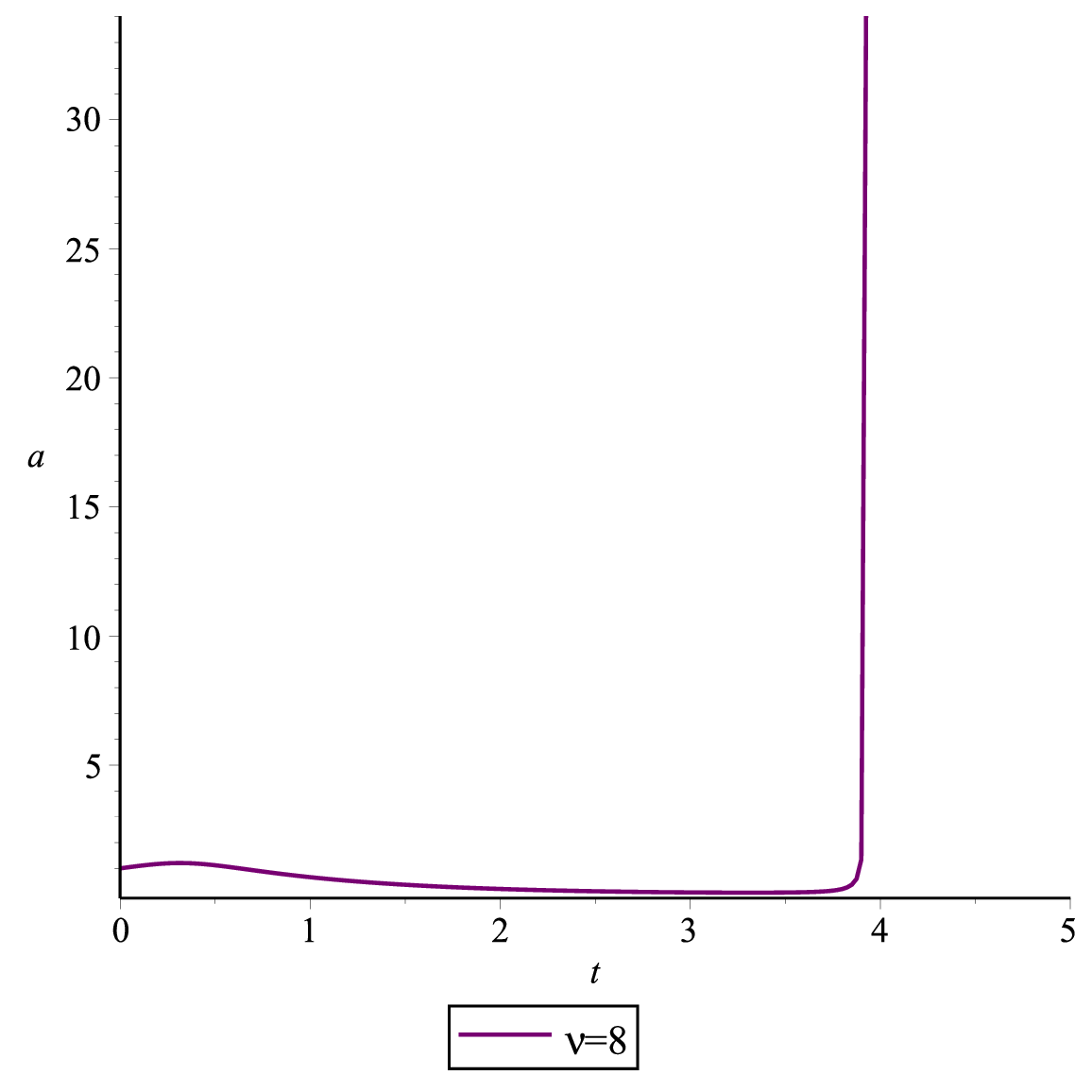}
	\caption{Variation of $a(t)$ for a large positive value of $\nu$.}
\label{f13}
   	\end{minipage}\hfill
	\begin{minipage}[t]{0.49\textwidth}
	\centering
   	\includegraphics[width=\linewidth]{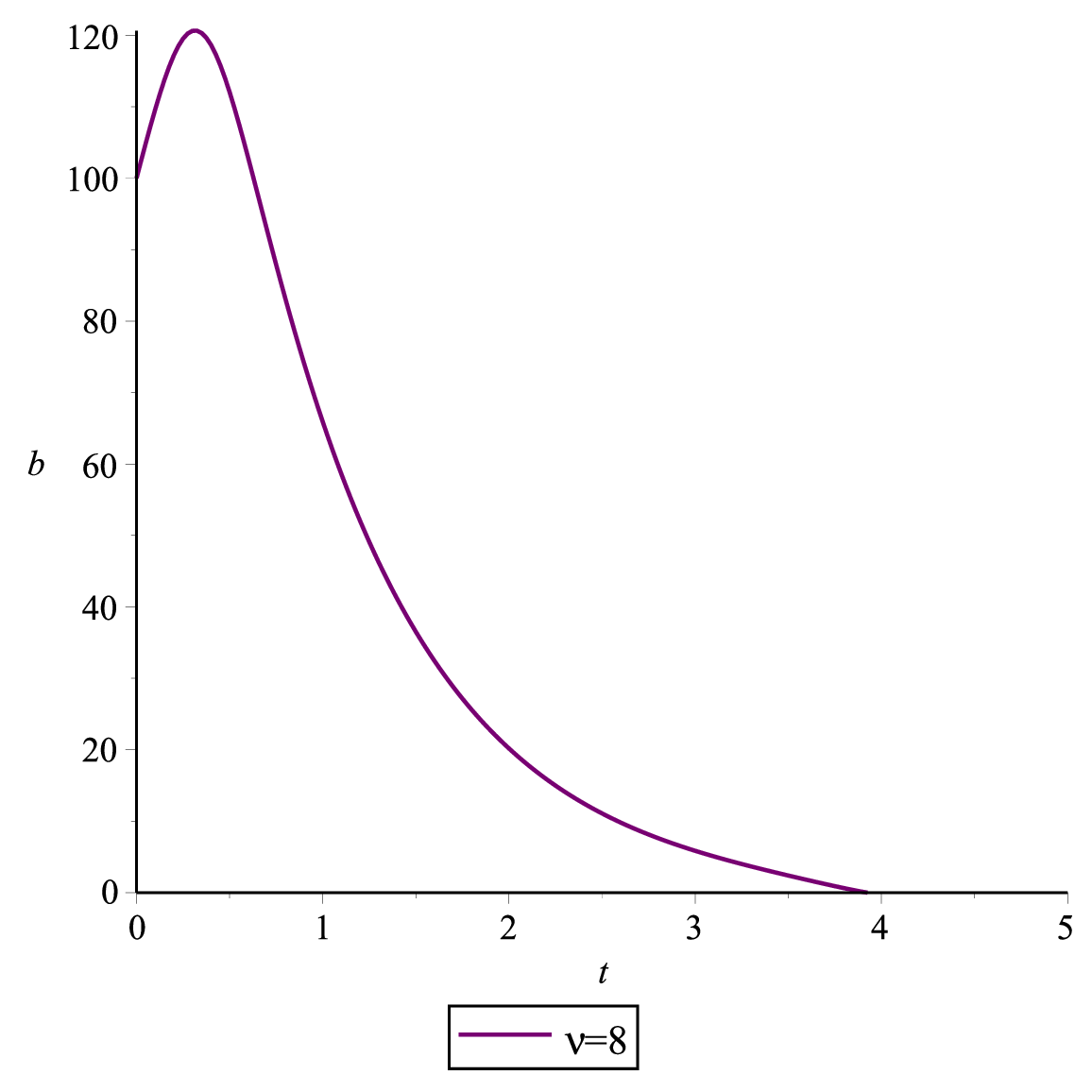}
	\caption{Variation of $b(t)$ for a large positive value of $\nu$.}
	\label{f14}
\end{minipage}\hfill
\end{figure}

\begin{figure}
\begin{minipage}[t]{0.49\textwidth}
		\centering
		\includegraphics[width=\linewidth]{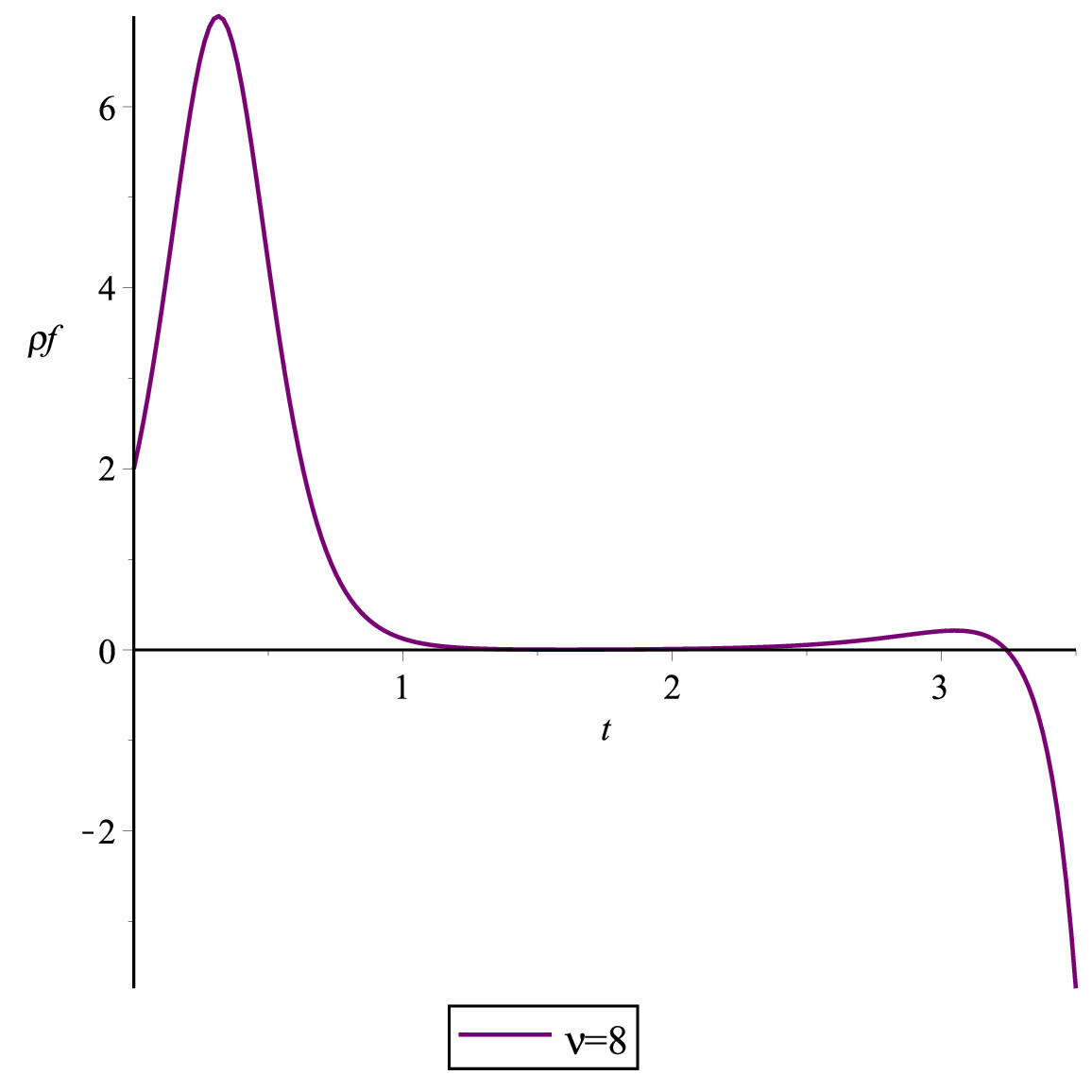}
		\caption{Variation of $\rho_f (t)$ for a large positive value of $\nu$.}
\label{f15}
\end{minipage}\hfill
\begin{minipage}[t]{0.49\textwidth}
		\centering
    	\includegraphics[width=\linewidth]{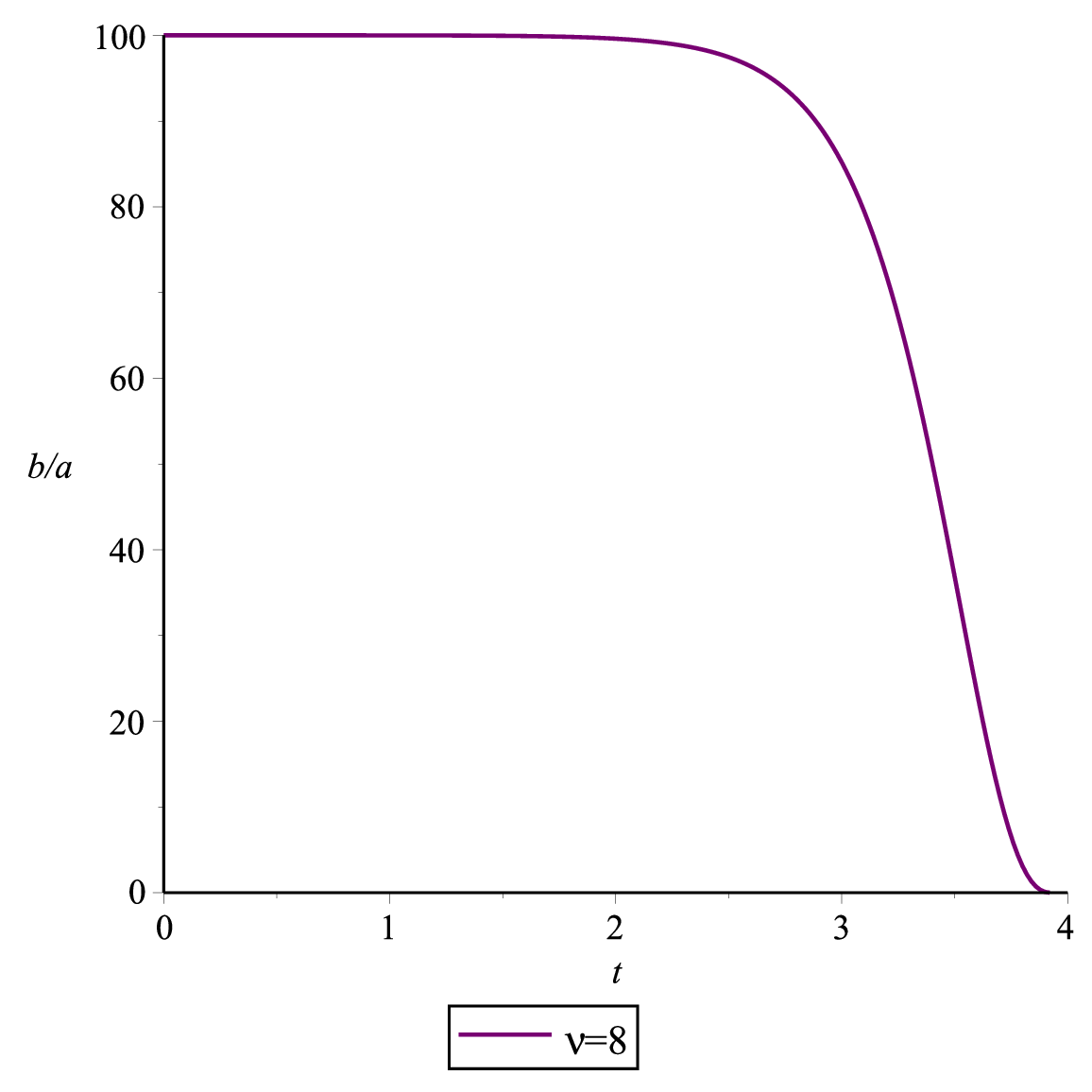}
		\caption{Variation of $b(t)/a(t)$ for a large positive value of $\nu$.}
\label{f16}
\end{minipage}\hfill
\end{figure}

On the basis of numerical analysis one notes that there may be a
value of $\nu$ where the tendencies related to the running stop
working. As an illustration, we show this situation in the four last
figures \ref{f13}, \ref{f14}, \ref{f15} and \ref{f16}, corresponding
to $\rho^0_{\Lambda} = 1$ and a huge unphysical value $\nu = 8$.
In this case, the term with $\nu$, in the Friedmann equation
(\ref{FriedmannI}) dominates over the basic term $\rho^0_\La$.
We included these plots just to illustrate the general situation that
may happen in the region of ``quantum dominance'' where the
running becomes very strong.

In Fig. \ref{f13} the scale factor $a(t)$ expands slowly and then
the Universe starts to contract. When the time gets close to
$t = 4$, an exponential expansion starts and goes until it abruptly
stops the expansion due to a final singularity at $t \approx 3.93$.
Fig. \ref{f14} shows the behavior of the second scale factor $b(t)$,
which also slowly expands until it reaches a maximum value at
approximately $t = 0.5$. After that, a contraction begins until it
reaches zero value and gives rise to a singularity, similar to a big
crunch, at the same value of time $t \approx 3.93$. Fig. \ref{f15}
demonstrates $\rho(t)$ with the same values of parameters. It is
easy to see that we meet (quite naturally) a singularity in the same
point.  To complete this part, in Fig. \ref{f16} one can observe
that the ratio $b(t) / a(t)$ remains approximately constant and then
begins to decrease at some point. This stage lasts until the same
point $t \approx 3.93$, when $b(t)$ goes to zero.

\section{Conclusions}
\label{secConc}

The most important qualitative result of our work is that, different
from the cosmic perturbations \cite{CCwave}, small values of the
phenomenological parameter $\nu$ do not affect the dynamics of
the anisotropic conformal factors, at least in the framework of
the KS-metric model \cite{KSori}. Taking into account the bound
for $\nu$ derived from the perturbations, one could conclude
that the possible running of the cosmological constant is irrelevant
for the dynamics of  anisotropic parameters, but this would be a
misleading statement. The reason is that in the very early Universe
the metric becomes isotropic very fast and, therefore, the two
kinds of deviation from the homogeneous and isotropic cosmology
occur at the distinct epochs. And, in the very early Universe we can
assume that the value of $\nu$ do not satisfy the aforementioned
bound. Assuming that this parameter is of the order one, we can
see how the running of $\La$ affect the isotropization.

Concerning anisotropic model and the role of the running in
isotropization of the metric, we can see that the model is tending
towards an isotropic configuration in the course of evolution, for
all values of $\nu$ which were considered. Also, one notes that
for smaller values of $\nu$, the ratio $b(t) / a(t)$ tends to a constant
value quicker. On top of this, the aforementioned constant value is
greater for smaller values of $\nu$. It is worth noting that since
both conformal factors depend only on time, these results do not
depend on the choice of coordinates. Another conclusion one can
draw from the plots in
Figures \ref{f1} - \ref{f12} is that the smaller the value of $\nu$,
the faster the expansion of the scale factors $a(t)$ and $b(t)$.
Furthermore, independent on the isotropy, the fluid density
$\rho_f$ goes to zero faster for smaller values of $\nu$. Both
tendencies hold for both positive and negative values of
the parameter $\nu$.
			
Finally, we conclude that the running of the cosmological constant
in the model with energy exchange between vacuum and matter
(radiation, in our case) sectors describes the accelerated expansion
and, for a moderate value of the phenomenological parameter $\nu$,
does not contradict very fast isotropization of the initially anisotropic
model.

The last observations concerns the possible extensions and
continuations of this work. Regardless the KS metric results look
convincing, it would be interesting to perform the same, or maybe
more detailed analysis for the Bianchi type I, or even more general,
metric. On the other hand, since the anisotropy under discussion
concerns only very early universe, when typical energies are
extremely high, it would be certainly interesting to include the
consideration of the effects of higher derivative terms, starting
from $R^2$. We hope to address this issue in a future work.

\section*{Acknowledgements}

Authors are grateful to A. Kamenshchik for useful correspondence.
I. Sh. acknowledges partial support from Conselho Nacional de
Desenvolvimento Cient\'{i}fico e Tecnol\'{o}gico - CNPq
under the grant 303635/2018-5. Vin\'{\i}cius G. Oliveira thanks
Coordena\c{c}\~{a}o de Aperfei\c{c}oamento de Pessoal de N\'{i}vel
Superior (CAPES) for his scholarship.
	



\newpage

\end{document}